# Developing a Collaborative and Autonomous Training and Learning Environment for Hybrid Wireless Networks

José Eduardo M. Lobo[1], Jorge Luis Risco Becerra[2], Matthias R. Brust[3], Steffen Rothkugel[4], Christian Adriano[5]

**Abstract** — *With larger memory capacities and the ability to link into wireless networks, more and more students uses palmtop and handheld computers for learning activities. However, existing software for web-based learning does not deal with new technological constraints that mobile devices bring with. Therefore, a new generation of applications for the learning domain that works on these restrictive mobile environments has to be developed. For this purpose, we introduce CARLA, a cooperative learning system that is designed to act in hybrid wireless networks. As a cooperative environment, CARLA aims to disseminate teaching material, notes, and even components of itself through both, the fixed and mobile network to find interested nodes. Due to the mobility of nodes, CARLA deals with upcoming problems as network partitions and synchronization of teaching material, resource dependencies, and time constraints.*

## INTRODUCTION

In recent years, mobile devices as cellular phones, notebooks, MP3-players, digital cameras, etc. became more commonplace in our daily life. In addition to *being mobile*, the desire of *being connected* with other devices or, in particular, the Internet, arises immediately.

More and more students take advantage of this progress to advance in their studies faster and more efficient. They use palmtop and handheld computers at home as well as at university to manage their homework or research. Technological advances like larger memory capacities, a variety of data input devices, Wi-Fi and Bluetooth for local ad-hoc communication as well as UMTS and GSM for linking to a backbone network drive progress in this respect.

The applicability and usefulness of pure ad-hoc networks is generally limited, due to the volatile nature of the environment. Devices are mobile, entering and leaving each others coverage area frequently. The density of devices is also subject to change. If the density drops below a certain critical threshold, communication might even become impossible. The envisioned hybrid wireless network, uses as "joker" a UMTS or GSM link to a fixed network with high deliver guaranties. However, these link causes additional costs and that use had to be avoided where is possible. Thus, the envisioned network fluctuates between a cost-free guarantee-less and expensive quality guarantying standard network. To deal with these kinds of hybrid wireless networks, the *Injection Communication* is proposed by two of the authors [5]

In terms of the software, there is a gap between these technological advances and the needs of the students. Several years ago researchers were faced with the challenge of realizing learning environments as an Internet-based server-client application. Since the use of mobile systems is increasing, the focus has shifted to the development of learning environments that fulfills both, technical requirements and restrictions of mobile systems and individual needs to support learning for the student.

This paper aims to investigate and develop a learning tool and a learning environment that fulfills these demands. It shows limitations of the learning domain using wireless link technologies, but also possibilities for new features.

The methodology for our investigations is twofold. First, we use our technical and pedagogical experiences in implementing web-based learning environments to figure out features for the envisioned software. Second, we use the scenario method to approach the issue. As result, we propose CARLA, a cooperative learning system for mobile environments working with hybrid wireless networks.

The next section describes the used methodology in detail. In section "Web-based Learning Systems", we analyze the web-based tutoring system AnITA2 and the web-based annotation system CALM to focus on essential learning resources and key factors explicitly described in section "Key Factors and Driving Forces". On that base, section "Scenario CARLA" describes the envisioned system named CARLA. The summery section closes this paper.

## METHODOLOGY

The methodology for our investigations is twofold. First, we use our technical and pedagogical experiences in implementing web-based learning environments to figure out features for the envisioned software. Second, we use the scenario method to create a promising scenario.

[1] Lobo, José Eduardo M., University of São Paulo, Brazil, Department of Computer and Digital Systems Engineering, jemlobo@aol.com
[2] Becerra, Jorge Luis Risco, University of São Paulo, Brazil, Department of Computer and Digital Systems Engineering, jorge.becerra@poli.usp.br.
[3] Brust, Matthias R., University of Luxembourg, Luxembourg, mr_brust@yahoo.de
[4] Rothkugel, Steffen, University of Luxembourg, Luxembourg, Steffen.Rothkugel@uni.lu
[5] Adriano, Christian, State University of Campinas, Brazil, Department of Computer Engineering, Christian@cit.com.br

## Scenario Generation

To find a plausible application scenario, we chose scenario generation as one working method. Scenarios have usually been used in strategic development contexts where it is crucial to shape a working strategy for an uncertain future [6][7]. Scenario planning generates several orthogonal futures with indicators for how these develop. In this paper, we selected important issues from the scenario method for our investigations, briefly described here:

- Identifying key factors
- Searching for the driving forces behind the key factors
- Creating scenario matrix by combining forces
- Identifying indicators that tell the direction the setting is heading

Scenarios have been credited [6] with four primary advantages: *robust decision making*, *stretching mental models*, *enhancing corporate perception* and *energizing the management*. Although scenario methods have mainly been developed for the corporate management context, we believe that scenarios would be advantageous for a distributed multi-disciplinary research such as CARLA.

## Own Experiences

During the rise of web-based application that started in the late 90, our research groups were working on web-based learning environments. AnITA2 [1], a tutoring system was implemented to facilitate autonomous training and realizing tests by using the web-technology as communication medium. CALM [2] focused on facilitating cooperative and collaborative work between students and tutors, also using the web-technology.

In regard to the new underlying network technology, we abstract key factors and driving resources from AnITA2 and CALM and transform it to be applied in the proposed learning environment, named CARLA.

## WEB-BASED LEARNING SYSTEMS

In the sub-sequent section, AnITA2 and CALM are described more detailed while some aspects are focused, because they are important for CARLA's design justification.

### AnITA2: University of Trier, Germany (2002)

AnITA2, a generic tutoring system for use in a web-based environment was developed and implemented. To get the desired generic characteristics, it was essentially to split knowledge and questions from its semantic in the learning unit. For this, test paradigms were introduced as generic semantic for AnITA2. AnITA realizes four test paradigms (cf. Fig. 1). *Free selection* displays questions in the order they appear in input file. *Causal Links Selection* is sensitive to right and wrong answers.

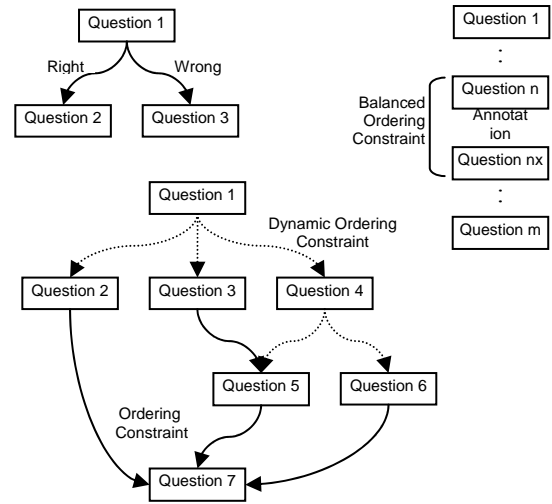

*Fig.1.* Causal Links, (Dynamic) Ordering Constraints and Balanced Ordering Constraints as Test-Paradigms in AnITA

*(Dynamic) Ordering Constrain Selection* follows a pre-determined order for selecting questions. A forced ordering constraint question can only be called upon other question as a reference. This paradigm takes into account the existence of a question that is a subpart of other questions and that according to the context cannot appear alone. A dynamic ordering constraint selection enables the system to choose randomly from different constraints. *Balanced Constraint Selection* is introduced with its values **n** and **p**. The **n** value implies an arithmetic average **a** for the last **n** selected questions. If **a** is greater than **p** the system may continue selecting the next question and following the ordering constraints, causal links or free selection. Otherwise, the selection repeats the last **n** questions.

AnITA2 is designed on client-server principles as a Web-application. AnITA2 uses component-based techniques for extensibility in respect to question types. Question types as *multiple-choice*, *fill-in questions* are already implemented.

### CALM: State University of Campinas, Brazil (2000)

We developed an annotation tool as part of a computer-based educational system, named CALM [1]. We had an evolutionary approach of adding new functionalities and exploring new uses for the annotation technology. We envisaged three scenarios, in increasing level of complexity: on-line asynchronous discussion, collaborative text authoring, and collaborative evaluation. The first and simpler was the support for discussion over a prepared hypertext available in the Web. The idea was to have students annotating the document and discussing about it. Annotations would be provided with symbols to denote agreement, disagreement, new facts, and issues. The second scenario was to support collaborative document edition by means of annotating and incorporating annotations to the final document. It made use of a more complex workflow for voting and consolidating annotated material.

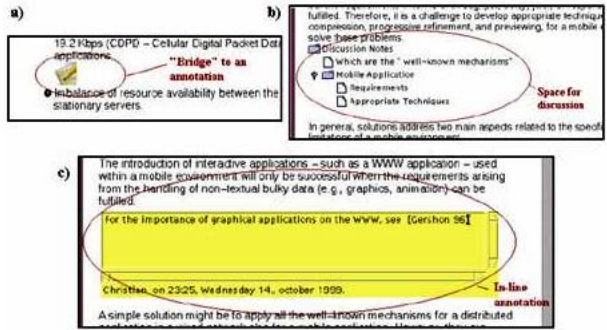

*Fig. 2.* Three annotation scenarios in CALM

The last scenario, which is again an evolution of the former ones, was to have a teacher and a student building an evaluation by means of exchanging annotations on the answers of an exam. The student would be doing the exam and exchanging points by receiving and doing annotations with the teacher. All these scenarios adopted three metaphors of annotation, demonstrated in Fig. 2, which are explained in detail by Adriano *et al.* [3].

## KEY FACTORS AND DRIVING FORCES

Implementation of novel learning scenarios frequently implies the adoption of new interaction paradigms provided by new media. Changes on media bring several issues to educational metaphors, such as deciding which characteristics, named here key factors and driving resources, should be maintained, removed and improved [4].

### Key Factors

Maintaining key factors for learning:
- Autonomous training: Distribution of official courses
- On-line asynchronous discussion: Adding annotations
- Collaborative authoring: Evaluating and directing the student's learning process
- Collaborative evaluation: Use of annotation metaphor

Adding key factors for learning:
- Collaborative training: Creating of own courses
- Meta-evaluation: Evaluating usefulness of a resource

Improving key factors for learning:
- Individual performance-based testing → Individual and grouped testing based on performance and cooperation

Maintaining key factors for the system:
- Extensibility of question types: Adding new question types by adding new system components
- Selection of question based on course: Describing the training by using the test paradigms

### Driving Resources

Identifying driving resources:
- Teaching material (Slides, text, articles, etc.)
- Annotation (Learning resource, CALM)
- Question (Learning resource, AnITA2)
- Course (Learning resource, AnITA2)
- Links (Learning resource, CALM)
- Software components for question types (System resource, AnITA2)

### Placing AnITA2 and CALM in CARLA

By using the scenario method, we could identify key factors and driving resources from two already implemented systems. The results serve as checklist for CARLA.

The next section describes the details and the problems of CARLA. In Fig. 3 the driving forces are placed around teaching material like slides, books, articles etc. This arrangement elucidates that the driving forces of AnITA2 is a vertical component and the driving forces of CALM is a horizontal component of CARLA. This result implies that the selected two systems are orthogonal to each other and, thus, cover a mighty scenario space. In that connection emerging problem zones between the key factors are discussed in the next section.

## SCENARIO CARLA

CARLA is a distributed learning application designed for mobile devices equipped with wireless communication adapters.

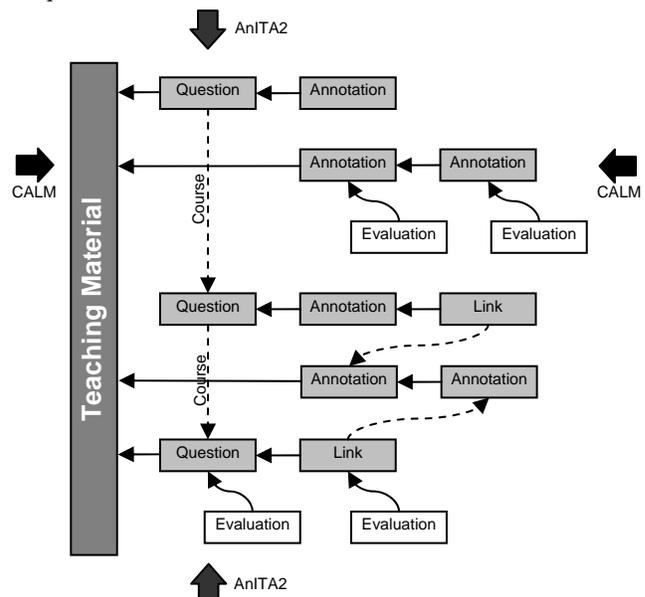

*Fig. 3.* AnITA2 as vertical component and CALM as horizontal component of CARLA shown in a possible local state of CARLA

*Teaching material* Students can use the system for example during and after lectures, being able to join cliques by sharing their material, and can help each other in a cooperative and collaborative way, e.g. to prepare exams. CARLA can primarily be used by students to manage teaching material like lecture slides, articles, papers, and glossaries. The teaching staff, i.e. professors, teachers, and tutors, uses the system to distribute such material among their students. Initially, all students basically have access to the same material. Possible locations of initial releases could be lecture halls and staff offices. Some students, however, might only have received subsets of the material released, for instance because of not having attended to a lecture. CARLA explicitly enables them to capture missing parts from their fellows later or from the backbone network, if the student wants to pay. During a lecture, a student's device can support him by showing the appropriate slides.

CARLA allows students to have personalized sets of their teaching material, as illustrated in Fig. 3, including annotations, questions, and links. During runtime, the subset of available material can be augmented by meeting other students.

*Questions* While recapitulating the material later on, students can use questions to get a deeper understanding of the topics covered. The questions can be provided by the staff or by other students.

*Annotations* Students are encouraged to add annotations to the teaching material, slides, and questions.

*Links* Moreover, students might discover additional relationships between some sections of the teaching material, their annotations, and the questions, adding them to their material. In CARLA, such relationships are called *links*.

*Evaluation* To prevent misleading or false additions to the teaching material from being distributed, each student is encouraged to evaluate such material received by others. This allows CARLA to detect and remove fakes, trying to improve the usefulness of the data.

*Courses* Together with initial questions the teacher may distribute a course that describes a pre-defined behavior for training or testing. As mentioned, question can be created instantly, also courses can be created and modified, assumed that the student has permission to modify.

### Forcing Cooperation

Cooperation must be forced by creating several goals for the student. Devoid of this step student could take advantage of existing information without give something in return, thus, causing the system' didactical breakdown.

One possibility here is to give an additional evaluation help to the tutor by introducing points for cooperative behavior. While the student actively participates on creating annotations, links, questions etc., points are accumulated. To avoid a misleading the points also depends on the numeric meta-evaluation of the added material by other students.

This method forces cooperation on the application level by introducing points for adding links, questions, and annotations. These cooperative-based points can be used together with knowledge-based points to evaluate the student in different levels of ability. It is even possible to automate the evaluation process, since all activities are known and the meta-evaluation principally exclude selfish use of the system by putting non-related data into the system.

### Collaborative authoring

Besides improving the learning potential in general, CARLA also fosters the work of the teaching staff. Influencing the learning process is possible at any time by adding additional links and/or annotations. Furthermore, CARLA allows redesigning the initial teaching material based on the students' feedback given by links, questions and annotations, thereby increasing its scope and usefulness.

### Dependencies of learning resources

Question, annotations, links, and evaluations are connected to at least one position in the initial teaching material or to another learning resource as described for instance in the course (see Fig. 3). A question type, e.g. multiple-choice, can just be displayed when the corresponded software component is loaded and initiated. The use of training and testing create high dependencies. The course must get following questions by using Causal Links. Ordering Constraints cause even higher dependencies that are difficult to tackle. The main point here is, if an Ordering Constrain can also be executed, even if not all questions of the Constrain are available?

In all, there are dependencies between the learning resources, teaching material, and software components for question types. To reduce costs, it is advisable in hybrid wireless networks to distribute learning resources and system components in an ad-hoc fashion. However, the existing dependencies cause problems. This behavior easily causes deadlocks, where for instance a link is submitted, but the destination or the source of the link is missing.

What information has to be sent first? As for instance links in the first device can depend of questions that maybe do not exist in the second device up to now, an information matching service must be initiated. With the result of the information matching service, it is possible do organize information exchanging in an effective way. If a deadlock is detected the backbone service can initiate a communication of an important resource in the local ad-hoc network.

### Injection Communication

AnITA2 and CALM are based on the client-server communication in a full infrastructured network. CARLA has to deal with infrastructureless networks, thus, the client-server model is assumed not to work for CARLA.

The whole CARLA scenario, including the game aspect, is expected to work much more efficient in hybrid wireless networks with the backbone joker than pure ad-hoc networks. However, a permanent backbone link is up to now very expensive. Two of the authors proposed a solution for this dilemma, introducing *Injection Communication* [5].

Using the Injection Communication, the backbone link is just use for a short time in "helpfully" situations. We can imagine a large PowerPoint-presentation with a lot of slides.

For one student, it would cost too much to get these slides over the backbone link. However, joining to a group of wireless connected devices that are also declaring interest in these slides, costs could be shared. It is also possible that the clique finds consents and each participant gets some of the slides and shares it in the clique in a peer-to-peer fashion etc.

### Game/Quiz

To increase the students' motivation, the teaching staff can use CARLA to create a quiz where the students act as players getting points for answering questions correctly. Similar to well-known TV game shows, players can use jokers for receiving hints. Three different kinds of jokers are supported: *link jokers*, *annotation jokers*, and *statistics jokers*. Link jokers allow exposing links pointing to additional material. Annotation jokers reveal annotations for a question, and statistics jokers can be used to show statistics indicating how other players answered this particular question, based on the available information. At a predefined time, e.g. at the last lecture of a course, the player with the highest rank wins. The use of wildcards is not restricted, but as soon as a wildcard is called on, the point number is halved. If another wildcard is called on, the point number is halved again and so on.

Considering the supposed wireless technology to create network communication, two important question rise up: When is a game finished? How to compare the results?

Ad-hoc networks that vary in density and mobility are extremely hard manageable for applications with time constraints. In this regard, during the application design we tried to avoid situations that require any hard time constraints. In spite of this, a challenging task might be defining the end of a quiz round to determine the champion.

In our case, where we have the option to use a backbone link to a fixed network in emerging cases, it is appropriate to terminate the game by inject each participating device, getting status information and transmit a ranking list. The time of interaction with the backbone link compared with the benefit it brings for the system, is very short.

### Some technical problems to tackle

Managing the teaching material together with the data added by the students is a challenging task. All devices participating, both of staff and students, form multiple ad-hoc network partitions over time. The temptation to use the available backbone link might be high to solve these problems, but one has not to forget the costs and other advantages when using the peer-to-peer communication. Nevertheless, the teaching material and the material added by the students like questions, annotations and evaluations needs to be shared among all CARLA users.

One critical problem is to decide which data has to be forwarded to which nearby device. To reduce network load, only devices interested in the same kind of knowledge will try to synchronize their teaching material. Another task is to keep questions with good evaluation quotes alive whereas questions with bad evaluation quotes are deleted automatically because their time to live is not updated and expires. This could minimize data overload in the network and maximize the benefits when getting data.

## SUMMARY

We analyze the web-based tutoring system AnITA2 and the web-based annotation system CALM to focus on essential learning resources and key factors and to create an application scenario based on hybrid wireless networks. The applied methods lead to the CARLA scenario that points out functionality change, communication and learning paradigm alteration that occur when the underlying network assumptions are changed. To deal with the communication model dilemma, we propose Injection Communication as underlying communication model.

Certainly, pure infrastructured settings aim at situations where users normally do not have access to any mobile device. The tendency, however, is that the use and accessibility of mobile devices providing services in the described manner is significantly increasing. Thus, for the near future the proposed combination between infrastructured and ad-hoc networking is going to become more and more relevant.